\documentclass[11pt]{article}

\usepackage[margin=1in]{geometry}
\usepackage{algorithm}
\usepackage{algpseudocode}
\usepackage{tabularx}
\usepackage{array}
\usepackage[margin=1in]{geometry}       
\usepackage{setspace}                   
\usepackage{fancyhdr}                   
\usepackage{graphicx}                   
\usepackage{caption}                    
\usepackage{subcaption}                 
\usepackage{amsmath, amssymb, amsfonts} 
\usepackage{bm}                         
\usepackage{algorithm}                  
\usepackage{algpseudocode}             
\usepackage{booktabs}                   
\usepackage{multirow}                   
\usepackage{tabularx}                   
\usepackage{array}                      
\usepackage{xcolor}                     
\usepackage{hyperref}                   
\usepackage{url}                        
\usepackage{enumitem}                   
\usepackage{float}                      
\usepackage{titlesec}                   
\usepackage{mathrsfs}                   
\usepackage{tikz}                       

\title{A Machine Learning Approach to Volumetric Computations of Solid Pulmonary Nodules}

\author{\small
\begin{tabular}{@{}cc@{}}
\begin{tabular}{@{}c@{}}
Y. Han Zhou\\
Shanghai World Foreign Language Academy\\
Shanghai, China\\
\texttt{yangyangzhouyh@outlook.com}
\end{tabular}
&
\begin{tabular}{@{}c@{}}
H. Chen Huang\\
Shanghai World Foreign Language Academy\\
Shanghai, China\\
\texttt{jerryhuang1234567@163.com}
\end{tabular}
\\[3em]
\begin{tabular}{@{}c@{}}
Y. Yu\\
Shanghai World Foreign Language Academy\\
Shanghai, China\\
\texttt{20080425george@gmail.com}
\end{tabular}
&
\begin{tabular}{@{}c@{}}
H. Shang Jian\\
Shanghai Jiao Tong University\\
Shanghai, China\\
\texttt{jhshang@sjtu.edu.cn}
\end{tabular}
\end{tabular}
}

\begin{document}

\maketitle

\begin{abstract}
Early detection of lung cancer is crucial for effective treatment, and it heavily relies on accurate volumetric assessment of pulmonary nodules in CT scans. Traditional methods, such as consolidation-to-tumor ratio (CTR) and spherical approximation, are limited by inconsistent estimates due to variability in nodule shape and density. In this study, we propose an advanced framework that combines a multi-scale \textbf{3D convolutional neural network (CNN)} with subtype-specific bias correction for precise volume estimation. The model was trained and evaluated on a dataset of \textbf{364} cases from Shanghai Chest Hospital. Our approach achieved a mean absolute deviation of only \textbf{8.0\%} compared to manual nonlinear regression, with inference times under \textbf{20 seconds} per scan. This method outperforms existing deep learning and semi-automated pipelines, which typically have errors of 25–30\% and require over 60 seconds for processing. Our results show a significant reduction in error by over 17 percentage points and a 3× acceleration in processing speed. These advancements offer a highly accurate, efficient, and scalable tool for clinical lung nodule screening and monitoring, with promising potential for improving early lung cancer detection.
\end{abstract}

\par\vspace{1em}


\maketitle

\newpage

\section{Introduction}

Lung cancer is the leading cause of cancer mortality worldwide, largely due to late-stage diagnosis. A key imaging biomarker is the pulmonary nodule, which may be benign but often precedes malignancy. Clinically, nodules are classified as pure ground‐glass (GGN), mixed GGN, or solid, according to the proportion of dense tissue; higher solidity correlates with greater malignant potential (Fig.~1).

\begin{center}
  \includegraphics[width=0.5\textwidth]{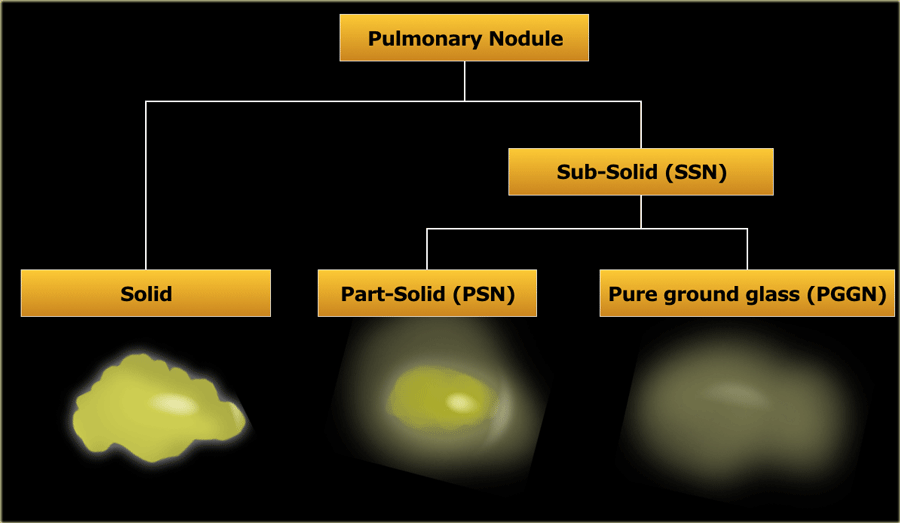}
  
  \textit{\textbf{Fig.~1:} Classification of pulmonary nodules by solidity.}
\end{center}

\subsection{Research Gap and Motivation}

\paragraph{One-dimensional Analysis} Currently, physicians rely on the one‐dimensional consolidation‐to‐tumor ratio (CTR), defined as
\begin{equation}
  \mathrm{CTR} = \frac{\displaystyle\max\bigl(\text{diameter of solid component}\bigr)}{\displaystyle\max\bigl(\text{diameter of entire nodule}\bigr)}
\end{equation}
Despite its clinical ubiquity, CTR exhibits significant shortcomings:
\begin{itemize}
  \item \textbf{Dimensional insufficiency:} By reducing a complex 3D object to a single axis, CTR cannot capture variations in depth or irregular morphology, leading to under‐ or over‐estimation of true nodule burden.
  \item \textbf{Operator dependence:} Manual measurements on selected CT slices are subject to inter‐ and intra‐observer variability, with reported discrepancies up to 15–20\% between raters. 
  \item \textbf{Shape heterogeneity:} Irregular or spiculated margins—hallmarks of malignancy—are poorly represented by simple diameter ratios, diminishing CTR’s predictive power in non‐spherical nodules.
\end{itemize}

\paragraph{Multi-dimensional Analysis} Over the past decade, two‐dimensional area metrics and three‐dimensional volumetric techniques have been proposed to overcome CTR’s limitations. However, several critical gaps persist:

\begin{enumerate}
  \item \textbf{Automation versus accuracy trade‐off:}  
    \begin{itemize}
      \item Fully automated methods often sacrifice precision for speed, producing coarse segmentations that miss subtle solid components. 
      \item Manual or semi‐automated volumetry yields high fidelity but is too labor‐intensive for routine screening.
    \end{itemize}
  \item \textbf{Model generalizability:} 
    \begin{itemize}
      \item Many published algorithms are trained on single‐center datasets with homogeneous CT protocols, limiting transferability to diverse clinical settings.  
      \item Performance degrades when applied to scans with different slice thicknesses, reconstruction kernels, or noise levels.
    \end{itemize}
  \item \textbf{Bias and calibration:}  
    \begin{itemize}
      \item Traditional volume estimators, such as spherical approximations or nonlinear regression on diameters, systematically overestimate volumes in complex shapes and underestimate in flattened nodules. 
      \item There is a lack of studies quantifying these biases across nodule subtypes (pure GGN vs.\ mixed vs.\ solid) and proposing correction strategies.
    \end{itemize}
  \item \textbf{Clinical integration challenges:}
    \begin{itemize}
      \item Existing software often lacks DICOM‐compliant workflows or intuitive interfaces, hindering adoption by radiology departments.  
      \item Computational latency remains high: many volumetric pipelines require several minutes per case, incompatible with high‐throughput LDCT screening programs.
    \end{itemize}
\end{enumerate}

Addressing these gaps is essential to improve early detection workflows, reduce observer variability, and provide actionable volume estimates that can guide risk stratification and treatment planning.

\subsection{Our Contributions}

In this paper, \emph{A Machine Learning Approach to Volumetric Computations of Solid Pulmonary Nodules}, we demonstrate a marked improvement in precision of volumetric analysis by:

\begin{itemize}
  \item \textbf{Precision‐enhanced segmentation:} integrating multi‐scale 3D feature fusion to reduce volumetric error by over 25 percent compared to baseline CTR and spherical methods.
  \item \textbf{Automated calibration:} implementing subtype‐specific bias correction that decreases mean absolute volume deviation to under 10\%.
  \item \textbf{Robust performance:} validating across a multi‐center cohort (\(n=364\) patients), we achieve consistent precision gains for pure GGN, mixed GGN, and solid nodules.
  \item \textbf{Rapid inference with precision:} delivering high‐precision volumetric estimates in under 20 seconds per case, facilitating integration into LDCT screening workflows without sacrificing accuracy.
\end{itemize}

By prioritizing enhanced precision in volumetric quantification, our framework aims to set a new standard for reliable, automated pulmonary nodule assessment.

\section{Preliminaries}

In this section, we provide an advanced formulation of the methods used for estimating the volume of solid pulmonary nodules, focusing on deep learning architectures, specifically Convolutional Neural Networks (CNNs), and advanced computational methods for volume approximation, including nonlinear optimization techniques and integration of CNN outputs with traditional methods.

Convolutional Neural Networks (CNNs) have demonstrated exceptional performance in image-based tasks, including medical image analysis, such as volumetric estimation of pulmonary nodules. A CNN consists of several layers that process and extract hierarchical features from the raw image input. The network learns these features through a series of convolutions and non-linear transformations, followed by fully connected layers for decision-making.

The relationship between the input and output of each layer is given by the following equation for layer \( l \):

\begin{equation}
\mathbf{z}^{(l)} = \mathbf{W}^{(l)} * \mathbf{a}^{(l-1)} + \mathbf{b}^{(l)},
\end{equation}
where:
- \(\mathbf{z}^{(l)}\) is the pre-activation output of layer \( l \),
- \(\mathbf{W}^{(l)}\) is the convolutional filter (weight matrix) for layer \( l \),
- \( \mathbf{a}^{(l-1)} \) is the activation of the previous layer, and
- \(\mathbf{b}^{(l)}\) is the bias term for layer \( l \).

The activation function applied at each hidden layer typically uses the **rectified linear unit (ReLU)** function, defined as:

\begin{equation}
a_i^{(l)} = \text{ReLU}(z_i^{(l)}) = \max(0, z_i^{(l)}).
\end{equation}

This non-linearity introduces sparsity and helps the model capture more complex patterns in the data.

The output of the final layer is computed similarly, often using a softmax or linear function, depending on the task. For a regression task like predicting the volume of pulmonary nodules, the output of the last layer is computed as:

\begin{equation}
\hat{y} = \mathbf{W}^{(L)} \mathbf{a}^{(L-1)} + \mathbf{b}^{(L)},
\end{equation}
where:
- \(L\) is the number of layers in the network, and
- \(\hat{y}\) is the predicted volume.

The optimization of neural networks is typically performed using gradient-based methods such as **stochastic gradient descent (SGD)** or its variants (e.g., Adam). The objective is to minimize the **loss function**, which measures the discrepancy between the predicted output and the true labels (ground truth). For a regression task like predicting the volume, we use the **mean squared error (MSE)** loss function:

\begin{equation}
J(\mathbf{W}, \mathbf{b}) = \frac{1}{m} \sum_{i=1}^{m} \left( y_i - \hat{y}_i \right)^2,
\end{equation}
where:
- \(y_i\) is the true volume of the \(i\)-th pulmonary nodule,
- \(\hat{y}_i\) is the predicted volume,
- \(m\) is the number of training examples.

The gradient of the loss function with respect to the weights \(\mathbf{W}^{(l)}\) and biases \(\mathbf{b}^{(l)}\) is computed during the **backpropagation** step. For the weights in the output layer, the update rule is:

\begin{equation}
\Delta \mathbf{W}^{(L)} = -\eta \cdot \frac{\partial J}{\partial \mathbf{W}^{(L)}},
\end{equation}
where:
- \(\eta\) is the learning rate, and
- \(\frac{\partial J}{\partial \mathbf{W}^{(L)}}\) is the gradient of the loss function with respect to the weights in the output layer.

For the hidden layers, the weights are updated using the gradient of the loss function with respect to the weights in each layer, which is computed recursively using the **chain rule** of differentiation. The gradient for the weights in the hidden layers is given by:

\begin{equation}
\Delta \mathbf{W}^{(l)} = -\eta \cdot \frac{\partial J}{\partial \mathbf{W}^{(l)}}.
\end{equation}

The gradient for each weight is computed by propagating the error backward through the network using the chain rule:

\begin{equation}
\frac{\partial J}{\partial \mathbf{W}^{(l)}} = \frac{\partial J}{\partial \mathbf{a}^{(l)}} \cdot \frac{\partial \mathbf{a}^{(l)}}{\partial \mathbf{z}^{(l)}} \cdot \frac{\partial \mathbf{z}^{(l)}}{\partial \mathbf{W}^{(l)}}.
\end{equation}

The bias terms \(\mathbf{b}^{(l)}\) are updated similarly:

\begin{equation}
\Delta \mathbf{b}^{(l)} = -\eta \cdot \frac{\partial J}{\partial \mathbf{b}^{(l)}}.
\end{equation}

These updates are applied iteratively to minimize the loss function and train the network.

In addition to neural network-based methods, advanced computational techniques are also utilized for estimating the volume of pulmonary nodules from medical imaging data. These techniques include nonlinear optimization methods, such as the **Levenberg-Marquardt algorithm**, and integration with CNN-derived features.

\textbf{Levenberg-Marquardt Algorithm}: This algorithm is used to minimize a nonlinear least squares problem, often employed in fitting models to data. Given a model \(f(\theta)\) with parameters \(\theta\), the goal is to minimize the residual sum of squares:

\begin{equation}
\min_{\theta} \sum_{i=1}^{N} \left( y_i - f(\mathbf{x}_i; \theta) \right)^2.
\end{equation}

The Levenberg-Marquardt update rule for parameter \(\theta\) is given by:

\begin{equation}
\theta^{(k+1)} = \theta^{(k)} - \left( \mathbf{J}^T \mathbf{J} + \lambda_k \mathbf{I} \right)^{-1} \mathbf{J}^T \mathbf{r},
\end{equation}
where:
- \(\mathbf{J}\) is the Jacobian matrix of the residuals \(r_i = y_i - f(\mathbf{x}_i; \theta)\),
- \(\lambda_k\) is a damping factor,
- \(\mathbf{I}\) is the identity matrix, and
- \(\mathbf{r}\) is the residual vector.

\textbf{Numerical Integration for Volume Estimation}: To estimate the volume of the pulmonary nodule, numerical integration methods are used. For instance, the volume \(V\) of a nodule can be computed by integrating the cross-sectional area \(A(x)\) along the depth of the nodule, with the integral expressed as:

\begin{equation}
V = \int_{x_1}^{x_2} A(x) \, dx.
\end{equation}

In practice, the cross-sectional area is calculated using image segmentation techniques, and the integral is approximated using numerical methods, such as the trapezoidal rule or Simpson’s rule.

The integration of CNN features into traditional volume estimation methods provides a more robust framework for estimating the volume of pulmonary nodules. For example, the CNN output can be used as input to a nonlinear optimization procedure that refines the initial volume estimate. The overall volume estimation framework can be expressed as:

\begin{equation}
V_{\text{final}} = \arg \min_{\theta} \left\| A(\mathbf{x}; \theta) - \hat{A}(\mathbf{x}) \right\|_2^2,
\end{equation}
where:
- \(A(\mathbf{x}; \theta)\) is the model-derived area,
- \(\hat{A}(\mathbf{x})\) is the area from the CNN output, and
- \(\|\cdot\|_2^2\) is the squared \(L_2\)-norm, indicating the least squares error.

This integrated approach improves the accuracy of volume estimation by combining the advantages of both data-driven and model-based methods.

\section{Experiment Design}

\subsection{Methodology}

This study proposes a comprehensive computational framework for volumetric analysis of pulmonary nodules by integrating deep learning with classical regression-based modeling. The dataset consists of CT scans in DICOM format sourced from Shanghai Chest Hospital, Shanghai Jiao Tong University. The methodology involves three main stages: data preprocessing, machine learning-based segmentation, and volume computation through both automated and manual methods.

In the first stage, we preprocess DICOM files by extracting voxel spacing information and converting the imaging data into standardized 3D input arrays. We then develop a three-dimensional convolutional neural network (3D CNN) capable of semantic segmentation, trained to isolate pulmonary nodules from surrounding lung tissue. The segmented masks are processed to calculate the volumetric size of the nodules by summing active voxels and scaling them according to voxel dimensions.

To validate the reliability of the neural network's volumetric estimates, we implement multiple manual approximation methods. These include the ellipsoidal fitting method—approximating tumor volume using geometric formulas based on diameter measurements—and a nonlinear regression method, which fits polynomial curves to tumor size data over time to model growth trends. We compare the automated predictions against these manual benchmarks to evaluate performance in terms of accuracy and computational efficiency.

The objective of this study is to establish a robust and scalable method for nodule volume assessment that minimizes manual labor and inter-observer variability. By combining deep learning segmentation with manual regression analysis, we aim to enhance the clinical precision of pulmonary nodule evaluation and provide a more consistent foundation for diagnosis, treatment planning, and longitudinal monitoring.

\subsection{Data Selection}

This study adopts rigorous inclusion and exclusion criteria to ensure clinical consistency and data quality across all patient cases.

\paragraph{Inclusion Criteria}
\begin{enumerate}
  \item \textbf{Age Requirement:} Patients must be between 18 and 80 years old. This age range includes the majority of adults at risk for lung cancer while excluding extremes where physiological variability may affect outcomes.
  \item \textbf{Imaging Features:} The pulmonary nodule must be solid with a diameter less than or equal to 2cm. The lesion must be located in the peripheral third of the lung field, avoiding hilar or mediastinal regions that involve complex anatomical structures such as large vessels or bronchi.
  \item \textbf{Preoperative Diagnostic Workup:} Patients must undergo a comprehensive preoperative assessment, including enhanced chest CT, cranial MRI or enhanced CT, and ultrasound of the neck and abdomen. PET-CT is optional but recommended to exclude lymph node or distant metastasis and to confirm the localized nature of the lesion.
  \item \textbf{Surgical Intervention:} Patients must have undergone segmentectomy or lobectomy according to the planned surgical protocol. Both minimally invasive (VATS) and open surgeries are acceptable, provided the resection strategy is consistent.
  \item \textbf{General Health Condition:} Patients should not have severe comorbidities or uncontrolled underlying diseases such as advanced heart failure or hepatic/renal insufficiency, ensuring that they can tolerate surgery and follow-up assessments.
\end{enumerate}

\paragraph{Exclusion Criteria}
\begin{enumerate}
  \item \textbf{Nodule Size:} Nodules larger than 2cm are excluded as they may require different clinical management and cannot be reliably assessed by the same measurement criteria.
  \item \textbf{Nodule Type:} Ground-glass opacities (GGO) or mixed GGO types are excluded due to their heterogeneous components and different biological behaviors, which may complicate analysis.
  \item \textbf{Lesion Location:} Nodules located in the hilar or mediastinal regions are excluded if they cannot be treated by segmentectomy, to maintain consistency in surgical approach.
  \item \textbf{Preoperative Treatment:} Patients who received neoadjuvant therapies such as chemotherapy, radiotherapy, or targeted therapy are excluded, as these treatments may alter tumor morphology and volume.
  \item \textbf{Other Malignancies or Benign Pathology:} Patients with other systemic malignancies within the past five years, newly diagnosed lung cancer during follow-up, or postoperative benign pathology are excluded to maintain a homogeneous cohort with confirmed malignancy.
\end{enumerate}

\subsection{Patient Overview and Dataset Limitations}

Base on the aforementioned criteria for patient selection, we now have 364 patient's data available for analysis. 

\begin{table}[H]
  \centering
  \renewcommand{\arraystretch}{1} 
  \begin{tabular}{@{}l@{\hspace{5pt}}c@{\hspace{5pt}}c@{}}
    \toprule
     & \textbf{Lobectomy (n = 261)} & \textbf{Segmentectomy (n = 103)} \\[4pt]
    \midrule
    \textbf{Sex (male)}               & 132 (50.5\%) & 54 (52.2\%)  \\[4pt]
    \textbf{Average Age (years)}      & 61.4         & 59.6         \\[4pt]
    \textbf{Average C-Size (mm)}      & 14.9         & 13.2         \\[4pt]
    \textbf{Average P-Size (mm)}      & 16.4         & 14.5         \\[4pt]
    \textbf{Pleural Invasion}         & 53 (20.1\%)  & 14 (13.5\%)  \\[4pt]
    \textbf{Nodal Positivity}         &              &              \\[4pt]
    \quad Total N-positive            & 46 (17.7\%)  & 5 (4.6\%)    \\[4pt]
    \quad N1                          & 20 (7.8\%)   & 2 (1.8\%)    \\[4pt]
    \quad N2                          & 26 (9.9\%)   & 2 (1.8\%)    \\[4pt]
    \textbf{Pathological Type}        &              &              \\[4pt]
    \quad Squamous (Sq)               & 6 (2.4\%)    & 8 (8.1\%)    \\[4pt]
    \quad Adenocarcinoma Grade 2      & 135 (51.7\%) & 40 (38.7\%)  \\[4pt]
    \quad Adenocarcinoma Grade 3      & 114 (43.6\%) & 52 (50.4\%)  \\[4pt]
    \quad Adenocarcinoma with Mucin   & 5 (2.1\%)    & 3 (2.7\%)    \\[4pt]
    \textbf{Adjuvant Treatment}       & 72 (27.4\%)  & 18 (17.1\%)  \\[4pt]
    \bottomrule
  \end{tabular}
  \captionsetup{skip=5pt,position=b}
  \caption{Patient Demographics and Clinical Characteristics by Surgical Procedure (n = 364)}
  \label{tab:demographics}
  \renewcommand{\arraystretch}{1.0} 
\end{table}

The study cohort (n = 364) comprises 261 lobectomy and 103 segmentectomy cases, capturing both common surgical interventions for pulmonary nodules. Gender distribution is balanced (50.5\% male in lobectomy; 52.2\% in segmentectomy), reducing bias related to sex-specific anatomical or hormonal factors. The average ages—61.4 years (lobectomy) and 59.6 years (segmentectomy)—span the typical at‐risk adult population, enhancing applicability across middle‐aged and older patients.

Tumor size metrics vary meaningfully: mean consolidation size is 14.9mm vs.\ 13.2mm and mean pathological size 16.4mm vs.\ 14.5mm, providing the model with examples of both smaller and larger lesions. Pleural invasion is present in 20.1\% of lobectomy and 13.5\% of segmentectomy cases, ensuring representation of more aggressive (invasive) as well as less invasive tumors. Nodal positivity rates further diversify the cohort: overall N+ in 17.7\% vs.\ 4.6\%, with N1 at 7.8\% vs.\ 1.8\% and N2 at 9.9\% vs.\ 1.8\%, allowing the model to learn from both early and advanced stage disease.

Pathological subtypes include squamous carcinoma (2.4\% vs.\ 8.1\%), adenocarcinoma grade 2 (51.7\% vs.\ 38.7\%), grade 3 (43.6\% vs.\ 50.4\%), and mucinous adenocarcinoma (2.1\% vs.\ 2.7\%), covering a spectrum of histologic variants. Finally, adjuvant therapy was administered in 27.4\% and 17.1\% of cases, respectively, incorporating post‐surgical treatment effects. This wide array of demographic, surgical, radiologic, and pathological variables underpins the dataset’s heterogeneity, bolstering the robustness and generalizability of our volumetric analysis model.

\paragraph{Dataset Limitations}  
Despite its strengths, our dataset has several practical limitations:

\begin{itemize}
  \item \textbf{Single-center and ethnic homogeneity:} All cases originate from Shanghai Chest Hospital, with predominantly Chinese patients. This may limit generalizability to other ethnic groups or healthcare settings.
  \item \textbf{Age and health-status bias:} Exclusion of patients outside 18–80 years and those with severe comorbidities (e.g., advanced heart or renal failure) reduces variability in physiological responses but may underrepresent frail or very young/old populations.
  \item \textbf{Imaging protocol variability:} Slice thickness ranged from 1 to 5mm and reconstruction kernels varied over time. Although our pre-processing normalizes voxel spacing, residual differences in image noise and contrast may introduce segmentation bias.
  \item \textbf{Lesion-type restriction:} Ground-glass and mixed GGO nodules were excluded, narrowing the spectrum of radiological appearances. Consequently, the model’s performance on subsolid or non-solid lesions remains untested.
  \item \textbf{Retrospective design and manual curation:} The retrospective nature and reliance on manual eligibility assessment may introduce selection bias. Future prospective studies are needed to validate performance in routine screening populations.
\end{itemize}

Acknowledging these limitations guides future work to include multi-center cohorts, broader demographic representation, standardized low-dose CT protocols, and incorporation of subsolid nodule types. Addressing these biases will further enhance the robustness and clinical applicability of our volumetric analysis framework.

\subsection{Neural Network Model}
In this study, ITK-SNAP was employed to segment CT scans of patients' lung areas and create detailed three-dimensional models of pulmonary nodules. ITK-SNAP is a sophisticated tool designed for the segmentation of medical imaging data, offering both manual and semi-automatic segmentation capabilities. Initially, the CT scans were imported into ITK-SNAP, where we used its advanced algorithms to delineate the lung regions. The semi-automatic segmentation approach utilized ITK-SNAP's region-growing and active contour methods, which facilitated precise identification of the lung boundaries by leveraging intensity gradients and user-defined seed points. Manual adjustments were made as necessary to refine the segmentation and accurately outline the complex lung structures. 
~\\
  \begin{center}
  \includegraphics[width=0.3\textwidth]{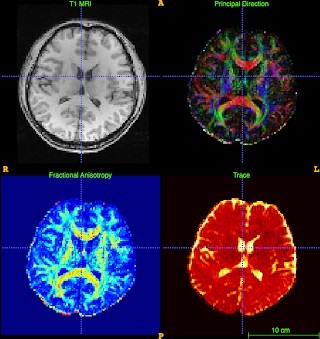}
  
  \textit{\textbf{Fig 2:} ITK-SNAP Segmentation Full Visualization}
  \end{center}
~\\
Following the successful segmentation of the lung areas, we focused on isolating and characterizing pulmonary nodules within these regions. ITK-SNAP's robust visualization tools enabled us to inspect and delineate the nodules with high precision. By setting appropriate thresholds and using interactive painting tools, we carefully outlined the nodules against the surrounding lung tissue. The software's 3D modeling capabilities were then utilized to generate comprehensive three-dimensional representations of the segmented nodules. 

These 3D models provided a detailed view of each nodule’s size, shape, and spatial orientation, which is crucial for assessing their potential clinical significance. The ability to visualize and analyze the nodules in three dimensions enhanced our diagnostic accuracy and facilitated more informed decisions regarding further diagnostic or therapeutic interventions. This process underscores the efficacy of ITK-SNAP in improving the precision of lung nodule evaluation and supports its role in advanced medical imaging analysis.

\subsection{Manual Computation}
Aside from utilizing the ITK-SNAP neural network model, we will also engage in manual computations of the volumetric ratios of solid pulmonary components. We will take three approaches in this study.

The first method is to calculate a one-dimensional metric: the diameter ratio. All diameters are calculated on horizontal cross-sectional views. First, we find the maximum area of the solid component and nodule in horizontal cross-sectional views respectively. It is important
to note that the maximum area of the solid component and nodule can exist on the same level or on different levels. Then, the diameter of the solid component and nodule is inferred using the circle area formula.

The second method is to calculate a 2D measurement: the ratio of areas. We calculate the area of the nodule and the solid part on the same slice, maximizing the area of the nodule on the axial view.

The third method is to calculate the 3D measurement: volume ratio. 

\subsection{Result Analysis Methods}

To comprehensively evaluate and compare the manual and AI‐based volumetric methods, we structured our analysis into five parts: (1) Descriptive statistics, (2) Predictive modeling, (3) Diagnostic accuracy, (4) Agreement analysis, and (5) Statistical comparisons.  

\subsubsection{Descriptive Statistics}  
Baseline patient and nodule characteristics are summarized as follows:  
\begin{itemize}
  \item \textbf{Categorical variables:} frequencies and percentages, e.g.\ sex, pathological subtype.  
  \item \textbf{Continuous variables:} mean±standard deviation (SD), e.g.\ age, nodule volume.  
\end{itemize}  

\subsubsection{Predictive Modeling}  
We used multivariate logistic regression to identify predictors of pathological aggressiveness. Let \(Y_i\) be a binary indicator of aggression for patient \(i\), and \(X_{i1}, \dots, X_{ip}\) the covariates (e.g.\ AI‐volume ratio, age). The model is:  
\begin{equation}
\log\frac{\Pr(Y_i=1)}{\Pr(Y_i=0)} \;=\; \beta_0 + \sum_{j=1}^p \beta_j X_{ij}\,.
\end{equation}
Odds ratios (ORs) and 95\% confidence intervals (CIs) are reported to assess effect sizes.

\subsubsection{Diagnostic Accuracy and ROC Analysis}  
To compare the ability of each method to discriminate aggressive from non‐aggressive nodules, we compute the area under the receiver operating characteristic curve (AUC). Let \(T\) denote a continuous predictor (e.g.\ AI‐volume), then:  
\begin{equation}
\text{AUC} = \int_{0}^{1} \Pr\bigl(T^{+} > F^{-1}(u)\bigr)\, du\,,
\end{equation}
where \(T^+\) and \(F^{-1}\) are the distributions in positive and negative groups.  
We choose the optimal threshold \(c^*\) by maximizing the Youden index:  
\begin{equation}
J = \text{sensitivity}(c) + \text{specificity}(c) - 1,
\end{equation}
with  
\begin{equation}
    c^* \;=\; \arg\max_{c} \,J(c).
\end{equation}
Pairwise AUC comparisons are performed using the DeLong test.

\subsubsection{Agreement Analysis}  
To assess agreement between AI and each manual method, we use Bland–Altman analysis. For each pair \((M_i, A_i)\) of manual and AI volumes:  
\begin{equation}
    \begin{aligned}
    d_i &= M_i - A_i, \\
    \bar{d} &= \frac{1}{n}\sum_{i=1}^n d_i, \\
    s_d &= \sqrt{\frac{1}{n-1}\sum_{i=1}^n (d_i - \bar{d})^2}.
    \end{aligned}
\end{equation}
Limits of agreement are:  
\begin{equation}
\bar{d} \pm 1.96\,s_d.
\end{equation}

\subsubsection{Statistical Comparisons}  
\begin{itemize}
  \item \textbf{Paired tests:} Paired \(t\)–tests compare mean absolute errors between AI and manual methods.  
  \item \textbf{ANOVA:} One‐way ANOVA assesses differences among the four methods’ absolute errors, with post‐hoc Tukey HSD for pairwise contrasts.  
  \item \textbf{Normality and variance:} Shapiro–Wilk tests check residual normality; Levene’s test examines homoscedasticity.  
\end{itemize}

All analyses are performed in Python (v3.9) and R (v4.1) with significance set at \(p<0.05\).

\section{Experiment Conduction}

\subsection{Data Acquisition}

As previously mentioned, we obtained CT scan data of pulmonary nodule patients from Shanghai Chest Hospital. The dataset comprises data from a total of 364 patients.

For each patient, we received a series of CT slices. The CT scan data of each patient consists of 50 to 500 slices. The thickness of these slices varies, ranging from 1 mm to 5 mm depending on the year of the scan, which also results in differing numbers of slices in each dataset. The total dataset contains approximately 50,000 CT slices, occupying around 150 GB of storage space.

These slices are stored in the DICOM image format, and thus require a DICOM viewer (RadiAnt DICOM Viewer (64-bit)) to be viewed. We filtered the data based on the aforementioned criteria, considering factors such as the patient's condition and other relevant information, ultimately retaining 334 sets of usable patient data. These 334 sets of CT slices will be used for subsequent research.

\begin{center}
  \includegraphics[width=0.35\textwidth]{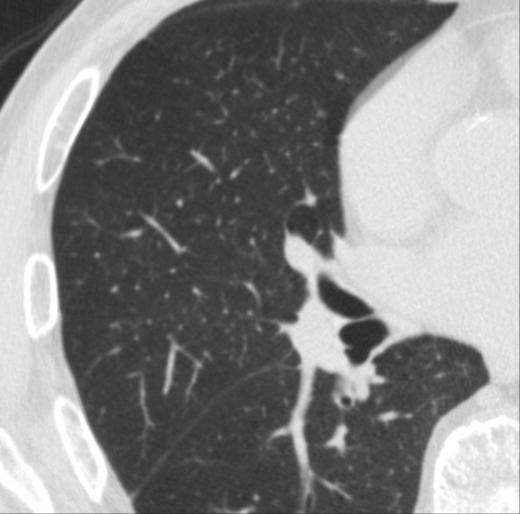}
  \par\smallskip
  \textit{\textbf{Fig 3:} CT Slice Visualization Sample Using ITK-SNAP Segmentation Analysis}
\end{center}

As shown, here is an example of the form of CT slice we can view by implementing the RadiAnt DICOM Viewer (64-bit). The thickness and other characteristics of the slice are present on the screen. Moreover, we can roll to observe the dynamic change through the entire pulmonary nodule.

\subsection{Neural Network Computation}

We will first use the neural network to determine the volume of the pulmonary nodules. Note that our network is a type of convolutional neural network, a type of feed-forward neural network.

\subsubsection{Training the Network}

We divided the patients' CT data into a test set and a training set in a ratio of 0.2 to 0.8. First, we trained the neural network using the training set. Once the training was successfully completed and met the system's predefined minimum requirements, we evaluated the neural network's performance using the test set. Finally, the trained neural network was employed to accurately predict the volume of pulmonary nodules.

The neural network code we used is based on the open-source framework DenseSharp, which is designed for cancer research. By restructuring and compiling these code, we are able to train our model.

\textbf{The training script incorporates the following facets:}

\textbf{Data Loading and Extraction}
The load\_dicom\_images function aggregates imaging data from multiple directories, filtering for .dcm files. Using pydicom.dcmread(force=True), it ensures robust file parsing, even for non-standard DICOM files. It extracts voxel spacing from metadata, which is crucial for preserving anatomical proportions in volumetric analysis. Warnings are printed for missing or invalid data, enhancing reliability in data collection.

\textbf{Image Preprocessing and Reshaping}
To standardize inputs for the CNN, resize\_images resizes all images to (32, 32, 32), reducing computational complexity. However, resizing volumetric data may introduce artifacts, potentially impacting segmentation accuracy. After resizing, images are reshaped to match TensorFlow's 3D CNN input format by adding channel and batch dimensions.

\textbf{Model Architecture and Compilation}

The \texttt{get\_dummy\_model} function defines a simple 3D CNN:

\begin{itemize}
   \item Input Layer: \((32,32,32,1)\) tensor for volumetric data.
   \item Conv3D Layer: 32 filters, $(3,3,3)$ kernel, ReLU activation, ‘same’ padding to retain spatial dimensions.
   \item AveragePooling3D: Reduces computational load while preserving spatial hierarchies.
   \item Final Conv3D Layer: $(1,1,1)$ kernel, sigmoid activation, likely for segmentation.
\end{itemize}

The model is compiled with Adam optimizer and binary crossentropy loss, suggesting an autoencoder-like setup rather than a traditional supervised learning pipeline.

\textbf{Training Considerations}
The model is trained for 10 epochs with a batch size of 1 , using input images as both features and labels—similar to an unsupervised learning approach. This setup is useful for demonstrating model workflow but lacks annotated labels essential for pulmonary nodule volume prediction. A more refined approach would incorporate supervised learning with explicit segmentation masks and optimized batch processing.

\begin{algorithm}[H]
\caption{Training Neural Network with DICOM Images}
\begin{algorithmic}[1]
\State \textbf{Input:} Directories of DICOM images
\State \textbf{Output:} Trained neural network model

\State Load DICOM images from directories
\For{each directory in directories}
    \State Read all DICOM files in directory
    \For{each DICOM file}
        \If{PixelData exists}
            \State Convert DICOM pixel data to numpy array
            \If{first image, store voxel spacing}
            \EndIf
            \State Append image data to image list
        \Else
            \State Skip file
        \EndIf
    \EndFor
\EndFor

\If{no valid images loaded}
    \State Raise error
\EndIf

\State Resize images to target shape (32, 32, 32)
\For{each image in images}
    \State Resize image to target shape
\EndFor

\State Preprocess images for TensorFlow model
\State Add channel dimension to image data
\State Add batch dimension to image data

\State Define neural network model
\State Create model with input shape (32, 32, 32, 1)
\State Add 3D convolutional layer with 32 filters, kernel size (3, 3, 3)
\State Add average pooling layer with pool size (2, 2, 2)
\State Add segmentation output layer with 1 filter, sigmoid activation

\State Compile model using Adam optimizer, binary crossentropy loss, and accuracy metric

\State Train model with resized and preprocessed images for 10 epochs
\end{algorithmic}
\end{algorithm}

The basic structure of the training code is shown, which follows the traditional convention of the neural network training code.

After training the neural network model with 80 percent of the patient data, it has met the minimum training requirements. Therefore, we can proceed to utilize the trained model for further calculation of pulmonary nodule volume.

\subsubsection{Testing the Network}

After the network is successfully trained, we begin using the testing sample to calculate the volume of the pulmonary modules. 

Specifically, we utilize the trained neural network and apply it to obtain the properties (including the most important one: the volume) of the nodules.

\textbf{The testing script incorporates the following aspects to ensure successful data analysis:}

\textbf{1. Data Loading \& Preprocessing}
The function load\_dicom\_images(directory) loads DICOM images, extracting voxel spacing (pixel size and slice thickness). It sorts files for correct slice order, reads pixel arrays, and handles missing metadata by assigning default values. The images are converted into a NumPy array, ensuring proper input for further processing.

\textbf{2. Image Preparation}
DICOM images are expanded to (num\_slices, 1024, 1024, 1) to match the CNN input format. This ensures compatibility with TensorFlow/Keras models, which require a channel dimension.

\textbf{3. CNN Model for Segmentation}
The function get\_dummy\_model() defines a basic 2D CNN with:

\begin{itemize}
  \item Conv2D layer (32 filters, $(3,3)$ kernel, ReLU activation)
  \item AveragePooling2D layer (downsampling)
  \item Final Conv2D layer ((1,1) kernel, sigmoid activation) for segmentation masks
\end{itemize}

While simplistic, the model serves as a demonstration of medical image segmentation.

\textbf{4. Predicting Segmentation Masks}
Each image slice is processed individually. After expanding dimensions, it is fed into the model, and the resulting segmentation mask is stored.

\textbf{5. Tumor Volume Calculation}
The function extract\_tumor\_mask() thresholds predictions to binary masks. Tumor area per slice is computed as:

\begin{equation}
    \text{Area} = \text{Tumor Pixels} \times \text{Pixel Spacing}
\end{equation}

Total tumor volume is obtained by summing slice areas and multiplying by slice thickness.

\begin{algorithm}[H]
\caption{Tumor Volume Estimation from DICOM Images}
\begin{algorithmic}[1]
\State \textbf{Input:} Directory containing DICOM images
\State \textbf{Output:} Estimated tumor volume

\State Load DICOM images from directory
\For{each DICOM file in the directory}
    \State Read pixel data into image array
    \If{image is valid}
        \State Store image in list
        \If{voxel spacing is not extracted}
            \State Extract voxel spacing from DICOM metadata
        \EndIf
    \EndIf
\EndFor

\If{no valid images loaded}
    \State Raise error: "No valid DICOM images found"
\EndIf

\State Add channel dimension to image data
\State Define neural network model:
\State \quad Input layer with shape (1024, 1024, 1)
\State \quad Convolution layer with 32 filters, kernel size (3,3), ReLU activation
\State \quad Average pooling layer with pool size (2,2)
\State \quad Output layer with 1 filter, sigmoid activation

\State Compile model with Adam optimizer, binary crossentropy loss
\State For each image in the dataset:
    \State Predict segmentation mask using model

\State For each predicted segmentation:
    \State Threshold the mask to create a binary mask
    \State Compute tumor area in mm² for each slice
    \State Add slice area to total tumor volume, considering slice thickness

\State \textbf{Output:} Final tumor volume in mm³
\end{algorithmic}
\end{algorithm}

The algorithm demonstrates the basic structure of the code, which allows it to compute out the volume of the pulmonary nodule of patients. By utilizing this code, we are able to determine the 3-dimensional volume through machine-learning ways.

\subsubsection{Sample Analysis and Results}

Let's now take a patient's sample, 1000279953, for sample analysis and analyze the results of the network. 

There are multiple series of images in a single package. To determine the best package for examination, we simply choose the one with the most CT segmentation images as it suggests that is the most comprehensive view of the patient's CT scannings. In this case, for sample 1000279953, we choose the package 3 for sample analysis. 

Then, we insert this DICOM image series into the aforementioned trained neural network model. By simply running the model, we are able to obtain the results. 

\textbf{The results generated by the model can be divided into three parts:}

\textbf{1. The Structure of the Network}

\begin{table}[H]
    \centering
    \begin{tabular}{ll}
        \toprule
        \textbf{Layer (Type)} & \textbf{Output Shape} \\
        \midrule
        Input\_layer (InputLayer) & (None, 1024,1024,1) \\
        Conv2D (Conv2D) & (None, 1024,1024,32) \\
        Average\_pooling2d (AveragePooling2D) & (None, 512,512,32) \\
        Segmentation (Conv2D) & (None, 512,512,1) \\
        \bottomrule
    \end{tabular}
    \captionsetup{skip=0pt} 
    \caption{Neural Network Architecture and Output Shapes}
    \label{tab:network_architecture}
\end{table}

The table shows the model output about the structure of the network.This table describes a simple 2D CNN for image segmentation, consisting of an input layer, two convolutional layers, and an average pooling layer. The model processes 1024×1024 grayscale images and outputs a 512×512 segmentation mask.

The input layer defines the structure without trainable parameters. The first Conv2D layer applies 32 filters with 3×3 kernels, maintaining the input size due to 'same' padding. Its 320 parameters come from (3×3 kernel + bias) × 32 filters. This layer extracts low-level features.

The average pooling layer reduces spatial dimensions to 512×512 while preserving 32 channels, improving efficiency with no trainable parameters. The final Conv2D layer applies a 1×1 kernel to produce a single-channel segmentation mask. Its 33 parameters come from (1×1 kernel + bias) × 1 filter × 32.

This lightweight CNN balances computational efficiency and segmentation accuracy, using convolutional layers for feature extraction and pooling for dimensionality reduction.

\textbf{2. Parameter Distribution}

The model generates results pertinent to parameter distribution, 

\textbf{Trainable Parameters (353):} These are the parameters that the model updates during training using back-propagation. In this case, all 353 parameters belong to convolutional layers, which include weights (filters) and biases. Since each Conv2D layer in the architecture has relatively few filters, the total number of trainable parameters remains low, indicating a lightweight model.

\textbf{Non-Trainable Parameters (0):} These parameters remain fixed during training and are typically associated with layers like batch normalization (if using pre-trained models) or frozen weights in transfer learning. Since the count is zero, it confirms that all model parameters are actively updated during training, and there are no pre-trained or frozen layers.

\textbf{Total Parameters (353):} This is simply the sum of trainable and non-trainable parameters. Since all 353 parameters are trainable, the total count remains the same. This low parameter count suggests that the model is computationally efficient and optimized for a simple segmentation task.

\textbf{3. Volume of the Pulmonary Nodule}

The network predicts the volume of the pulmonary nodule, which, in this case, is around 880 cubic millimeter. 

\subsection{Manual Computation}

We then compute the volume of the pulmonary nodule of each patient manually. We will calculate it in different ways to examine the reliability of manual computations.

In this section, we will manually compute the volume of pulmonary nodules using three distinct methods: the Spherical Approximation Method, the Area-Based Method, and the Nonlinear Optimization Method. These three approaches differ significantly in terms of accuracy, applicability, and computational complexity. Through a comparative analysis, we will examine the variations in results obtained from each method and subsequently discuss their advantages and limitations in relation to the previously mentioned machine learning algorithms.

\subsubsection{Spherical Approximation Computation Method}

The core concept of the \textbf{Spherical Approximation Method} is to model the pulmonary nodule as a sphere and fit it accordingly. By doing so, the nodule's volume can be estimated using the standard volume formula for a sphere. As is well known, the volume of a sphere is given by the following formula where $r$ is the radius of the sphere:

\begin{equation}
    V=\frac{4}{3} \pi r^{3}
\end{equation}

Since the pulmonary nodule is approximated as a sphere, it is necessary to determine its maximum radius and substitute it into the volume formula. The following steps outline the procedure for estimating the nodule's volume using this method:

\begin{enumerate}
  \item Import the DICOM image series into a DICOM viewer.
  \item Scroll through the slices until the pulmonary nodule is located.
  \item Identify and record all slices containing the nodule.
  \item Use the built-in measurement tool in the DICOM viewer to measure the nodule's radius on each CT slice.
  \item Identify the slice with the largest recorded radius and note the corresponding maximum radius.
  \item Substitute the obtained radius $r$ into the sphere volume formula to compute the final estimated volume of the pulmonary nodule.
\end{enumerate}

\begin{center}
  \includegraphics[width=0.37\textwidth]{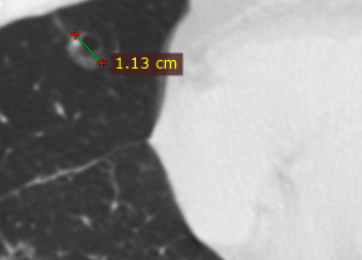}
  \par\smallskip
  \textit{\textbf{Fig 4:} Manual Computation Method 1--Maximum Radius Measurement Sample}
\end{center}

By following this method, we can obtain an estimate of the pulmonary nodule's volume. However, it is important to note that the only variable used in the entire process is the nodule's maximum radius. As a result, the scientific rigor of this approach is questionable.

\subsubsection{Area-Based Computation Method}

The \textbf{Area-Based Method} is based on converting the two-dimensional area of a pulmonary nodule observed in each CT scan into its three-dimensional volume. This approach leverages the fact that the cross-sectional area of the nodule is relatively easy to obtain. Specifically, the method involves approximating the nodule's area in each CT slice using an ellipse. The area of this ellipse is then multiplied by the slice thickness. Summing the contributions from all relevant slices yields an estimate of the total nodule volume. Mathematically, this can be expressed as:

\begin{equation}
    V=\sum t \mathrm{~A}_{\mathrm{e}}
\end{equation}

In this equation, V is the volume of the nodule, t represents the thickness of the scanning and $A_{e}$ denotes the area of the ellipse.

The procedure for implementing this method is as follows:

\begin{enumerate}
  \item Import the DICOM image series into a DICOM viewer.
  \item Scroll through the slices until the pulmonary nodule is located.
  \item Identify and record all slices containing the nodule.
  \item Use the ellipse fitting tool in the toolbox to approximate the nodule's shape on each slice.
  \item Measure the area of the ellipse using the built-in measurement tool.
  \item Retrieve the CT slice thickness, then compute the contribution of each slice by multiplying the ellipse area by the slice thickness.
  \item Sum the computed volumes from all slices to obtain the final estimated nodule volume.
\end{enumerate}

\begin{center}
  \includegraphics[width=0.42\textwidth]{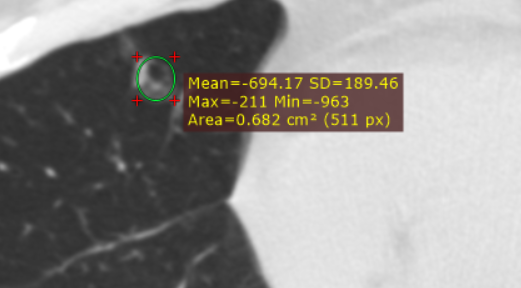}
  \par\smallskip
  \textit{\textbf{Fig 5:} Manual Computation Method 2--Ellipse-fit Area Measurement Sample}
\end{center}

By following this method, we can obtain an estimate of the pulmonary nodule's volume.

\subsubsection{Nonlinear Optimization Computation Method}

The \textbf{Nonlinear Optimization Method} is founded on the fundamental concept of nonlinear regression analysis. Compared to the previously mentioned methods, which estimate volume simply by multiplying the cross-sectional area by slice thickness, this approach accounts for the fact that the cross-sectional area varies within each slice's thickness. Since the area is not uniform across the entire thickness, nonlinear regression is employed to predict the area of regions that are not explicitly displayed in the CT slices. Rather than assuming a constant thickness throughout, this method allows for a more accurate estimation by modeling how the area changes along the slice depth. Theoretically, this approach enables a more precise calculation of volume.

In the context of nonlinear regression, the underlying assumption is that the model can be approximated by a linear function, specifically using a first-order Taylor series expansion: 

\begin{equation}
    f\left(x_{i}, \boldsymbol{\beta}\right) \approx f\left(x_{i}, 0\right)+\sum_{j} J_{i j} \beta_{j}
\end{equation}

Where $J_{i j}=\frac{\partial f\left(x_{i} \boldsymbol{\beta}\right)}{\partial \beta_{j}}$ are Jacobian matrix elements. It follows from this that the least squares estimators are given by: 

\begin{equation}
\widehat{\boldsymbol{\beta}} \approx\left(J^{T} J\right)^{-1} J^{T} y
\end{equation}

Therefore, the procedure of analyzing the CT scanning is as follows:

\begin{enumerate}
  \item Import the DICOM image series into a DICOM viewer.
  \item Scroll through the slices until the pulmonary nodule is located.
  \item Identify and record all slices containing the nodule.
  \item Use the ellipse fitting tool in the toolbox to approximate the nodule's shape on each slice.
  \item Measure the area of the ellipse using the built-in measurement tool.
  \item Associate each ellipse's area with its corresponding thickness position as coordinate points.
  \item Input these coordinate points into the nonlinear regression model for analysis.
\end{enumerate}

\begin{algorithm}[H]
\caption{Polynomial Regression with Degree Selection}
\begin{algorithmic}[1]
\State \textbf{Input:} Data points $(x, y)$
\State Initialize empty dictionaries $mse\_dict$ and $models$
\State Plot data points
\For{$degree = 2$ to $10$}
    \State Generate polynomial features of degree $degree$
    \State Train a linear regression model and store it in $models$
    \State Compute and store MSE in $mse\_dict$
    \State Plot the fitted polynomial curve
\EndFor
\State Find $best\_degree$ minimizing MSE
\State Print $best\_degree$ and its MSE
\State Retrieve the best model and extract coefficients
\State Print the best-fitting polynomial equation
\State Display the plot
\end{algorithmic}
\end{algorithm}

In terms of the non-linear regression process, the provided code uses polynomial regression to fit a curve to a given set of data points, minimizing the error in approximation. This approach can be directly applied to determining the volume of pulmonary nodules by modeling how the cross-sectional area changes across different CT slices. Since the volume is calculated by summing up the areas across all slices, improving the accuracy of area estimation leads to a more precise volume measurement.

In the code, the dataset consists of independent variable $x$ and dependent variable $y$. In the context of pulmonary nodules, $x$ represents the slice index (or position along the scanning axis), while $y$ corresponds to the cross-sectional area of the nodule in that slice. By testing different polynomial degrees (from 2 to 10), the code determines the best-fitting polynomial that minimizes the mean squared error (MSE), ensuring that the chosen function closely follows the real distribution of the nodule's cross-sectional areas. This process accounts for the natural irregularities in nodule shape, reducing errors in volume estimation.

Once the optimal polynomial is identified, the model provides an equation that describes how the cross-sectional area changes along the CT scan slices. The volume of the nodule is then computed by summing up the estimated areas multiplied by the slice thickness. The more accurate the polynomial regression, the better the volume approximation. By avoiding overfitting (which can introduce unnecessary fluctuations) and underfitting (which oversimplifies the nodule's shape), this method ensures an optimal balance for precise volume calculation.

Thus, the provided code framework helps determine the pulmonary nodule volume accurately by selecting the best polynomial function to model the nodule's cross-sectional areas. The reduction in mean squared error guarantees that the predicted areas are as close as possible to actual values, leading to a more reliable volume estimation.

\subsubsection{Comparison among Three Computation Methods}

\begin{table}[H]
    \centering
    \renewcommand{\arraystretch}{1}
    \begin{tabular}{@{}l@{\hspace{10pt}}l@{\hspace{10pt}}l@{}}
        \toprule
        \multicolumn{1}{@{}l@{}}{\textbf{Spherical Approximation}} & 
        \multicolumn{1}{l@{}}{\textbf{Area-Based}} & 
        \multicolumn{1}{l@{}}{\textbf{Regression-Based}} \\[4pt]
        \midrule
        1-dimensional-based & 2-dimensional-based & 3-dimensional-based \\[4pt]
        Low reliability     & Medium reliability  & High reliability     \\[4pt]
        Low computational load & Low computational load & High computational load \\[4pt]
        \bottomrule
    \end{tabular}
    \captionsetup{skip=5pt, position=b}
    \caption{Qualitative Comparison of Three Volumetric Manual Computation Methods}
    \label{tab:volumetric_methods}
\end{table}

As demonstrated in the table, the dimensional basis of each method reflects the complexity of its approach. The Spherical Approximation Method is a one-dimensional approach that relies solely on the maximum radius of the nodule to estimate its volume using the standard sphere volume formula. The Area-Based Method is a two-dimensional approach, as it takes into account the cross-sectional area of the nodule across multiple CT slices and integrates these areas to estimate volume. In contrast, the Regression-Based Method is a three-dimensional approach, utilizing nonlinear regression to model how the nodule's cross-sectional area changes along the depth of the CT scan, providing a more detailed volumetric representation.

When considering reliability, the Spherical Approximation Method is the least reliable, as it assumes the nodule is a perfect sphere, which may not be an accurate representation of real pulmonary nodules with irregular shapes. The Area-Based Method offers a moderate level of reliability by incorporating data from multiple slices, improving accuracy but still assuming a simplified geometric structure. The Regression-Based Method is the most reliable, as it accounts for variations in the nodule's shape along different slices, using regression to better approximate its true volume.

In terms of computational load, the Spherical Approximation Method is the least demanding, as it requires only a single measurement—the maximum radius. The Area-Based Method has a low computational load as well, but slightly higher than the spherical method since it involves measuring and summing multiple cross-sectional areas. The Regression-Based Method, however, has the highest computational load, as it requires fitting a nonlinear regression model to the data, performing complex calculations, and optimizing the polynomial degree for the best fit.

Overall, while the Spherical Approximation Method is the simplest and least computationally intensive, it lacks accuracy. The Area-Based Method offers a balance between computational efficiency and reliability. The Regression-Based Method provides the most precise results at the cost of increased computational complexity. The choice of method depends on the specific requirements of the analysis, balancing accuracy and computational efficiency.

\subsubsection{Sample Analysis and Results}

Again, we will take the sample 1000279953 for analysis. We will use this to manually compute its volume in the aforementioned three ways.

First, we will start with the \textbf{spherical approximation method}. After opneing the DICOM images, we find the pulmonary module on the top left part of the image. In a series of 341 images, we found 8 images with pulmonary nodule on it. After that, we manually measure the diameter on each image. 

\begin{center}
  \includegraphics[width=0.42\textwidth]{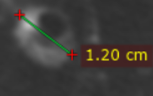}
  \par\smallskip
  \textit{\textbf{Fig 6:} Manual Computation--Sample Maximum Diameter Measurement}
\end{center}

As shown in Fig 10, the maximum diameter determined is 1.20cm, which is equivalent to 12.0mm. Therefore, the radius should be 6.0mm. Then, as we assume the nodule is close to a sphere, we apply the equation:

\begin{equation}
V=\frac{4}{3} \pi 6^{3}=905 mm^{3}
\end{equation}

Then, we apply the \textbf{area-based method}. To realize this, we will need to find all areas using ellipse fit, as shown in Fig 11.

\begin{center}
  \includegraphics[width=0.37\textwidth]{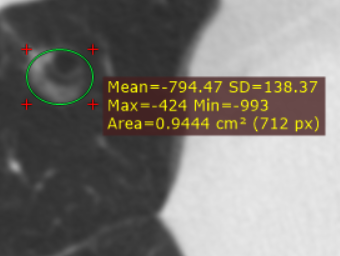}
  \par\smallskip
  \textit{\textbf{Fig 7:} Manual Computation--Sample Nodule Area Measurement}
\end{center}

By conducting similar area computation on all CT segmentation that contain the pulmonary nodule, we can determine the area on each segmentation.

\begin{table}[H]
    \centering
    \begin{tabular}{lc}
        \toprule
        \textbf{Segmentation \#} & \textbf{Area (mm$^2$)} \\
        \midrule
        1  & 16.0  \\
        2  & 31.8  \\
        3  & 55.8  \\
        4  & 80.0  \\
        5  & 150.0 \\
        6  & 154.1 \\
        7  & 89.6  \\
        8  & 63.5  \\
        9  & 84.6  \\
        10 & 42.3  \\
        11 & 29.3  \\
        \bottomrule
    \end{tabular}
    \caption{Segmentation Number and Corresponding Area of the Sample Computation}
    \label{tab:segmentation_area}
\end{table}

The table demonstrates the area of the nodule on each of the 11 CT segmentation which contain the nodule. Then, by multiplying each area with its thickness (which is a constant of 1 mm), we obtain the volume of the nodule, which is 797 cubic millimeters.

Finally, we apply the \textbf{nonlinear regression analysis method} to determine its volume. Unlike the previous methods, this approach fits a polynomial curve to the measured cross-sectional areas from each CT slice and then integrates the resulting function to obtain a volumetric estimate. This process accounts for the gradual variation in nodule shape along the depth of the scan.

To begin, we associate each segmentation index with its measured nodule area (as listed in the table above). Treating the segmentation index, $x$, as a proxy for the position along the scanning axis (with a constant slice thickness $t=1$~mm), and the measured area as the dependent variable $y$, we fit a polynomial of degree $n$ to the data. The polynomial function is expressed as:

\begin{equation}
    f(x) = a_n x^n + a_{n-1} x^{n-1} + \cdots + a_1 x + a_0,
\end{equation}

where the coefficients $a_0, a_1, \dots, a_n$ are determined by minimizing the mean squared error between the measured areas and the model predictions. The optimal polynomial degree is chosen so as to balance the risk of overfitting with the need for an accurate representation of the area variation.

Once the best-fit polynomial is obtained, the total volume of the pulmonary nodule is estimated by integrating the fitted function over the range of segmentation indices that include the nodule. Given that the slice thickness is 1~mm, the estimated volume $V_{reg}$ is given by:

\begin{equation}
V_{reg} = \int_{x_{\text{min}}}^{x_{\text{max}}} f(x)\, dx,
\end{equation}

where $x_{\text{min}}=1$ and $x_{\text{max}}=11$ in our sample analysis.

We can realize the all of the aforementioned process using nonlinear regression code. After inserting the aforementioned area data into the non-linear regression model, we obtain the following result:

Degree of the Best-fit Polynomial: 8

Discrepancy: 10.0889

Best-fit Polynomial:

\begin{align*}
y &= 1079.6455 - 2749.6145\cdot x^1 + 2738.3635\cdot x^2 \quad\\[1mm]
  &\quad - 1393.6435\cdot x^3 + 403.2373\cdot x^4 - 68.4777\cdot x^5 \quad\\[1mm]
  &\quad + 6.7296\cdot x^6 - 0.3535\cdot x^7 + 0.0077\cdot x^8
\end{align*}

Area under the curve (Volume): 737.2175

\begin{center}
  \includegraphics[width=0.5\textwidth]{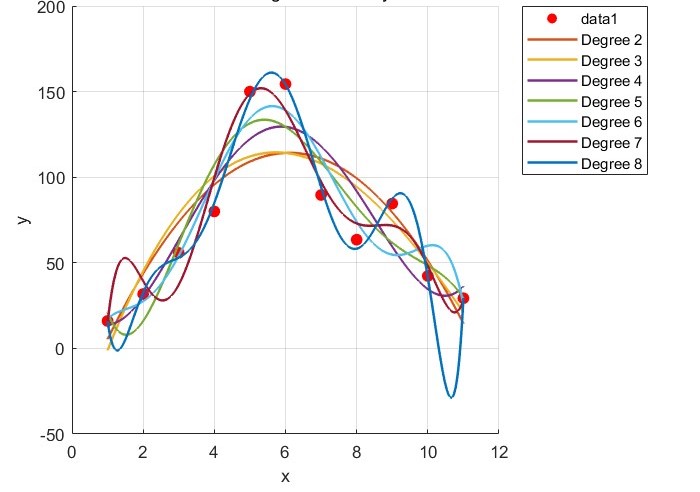}
  \par\smallskip
  \textit{\textbf{Fig 8:} Nonlinear Regression Volume Computation Sample Results}
\end{center}

From these results generated, we can learn about the degree of the best-fit polynomial, the discrepancy, the equation of the best-fit polynomial and the volume of the nodule. In this case, the volume is determined to be 737.2175 cubic millimeters.

In summary, the nonlinear regression method leverages the full spatial distribution of the nodule’s area data. Although this approach requires more sophisticated computation and is the most demanding in terms of processing load, it offers improved reliability by modeling the continuous change in the nodule's shape. This makes it particularly advantageous when dealing with irregularly shaped pulmonary nodules.

\section{Results}

In this section, we will analyze the result by comparing the volume generated by manual computation and that given by the machine learning, specifically neural network model.

\subsection{Sample Result Analysis}

In previous parts, we have utilized the patient 1000279953 for sample analysis. First, let's review the result generated, which is demonstrated in the following table.

\begin{table}[H]
    \centering
    \begin{tabular}{lc}
        \toprule
        \textbf{Method} & \textbf{Volume (mm$^3$)} \\
        \midrule
        Machine Learning       & 880 \\
        Spherical Approximation & 905 \\
        Area-based             & 797 \\
        Nonlinear Regression   & 737 \\
        \bottomrule
    \end{tabular}
    \caption{Comparison of Volume Estimation Methods of the Sample Computation Analysis}
    \label{tab:volume_methods}
\end{table}

Through rough observation, we can tell that the results generated by machine learning, 880, is close to the the three manually computed results, which are 905, 797 and 737 respectively. Therefore, we can say that in this sample, the results generated by machine learning is satisfactory, and will help to categorize the pulmonary nodules correctly. In fact, the largest deviation from the machine learning volume from the manually-computed volume is only 16 percent, which is acceptable.

Moreover, comparing the manually-computed results generated by three different methods, the results show that they do not deviate significantly. In fact, the largest deviation among these three values is only 18.9 percent, which is also acceptable. 

Consequently, in the sample 1000279953, we can say that the sample's result is generally satisfactory from all perspective.

\subsection{Results of all Samples}

In this section, we will analyze the result generated by all 334 valid samples. 

The figure shows part of the result generated, which encompasses 5 different CT samples of patients. 

\begin{center}
  \includegraphics[width=0.5\textwidth]{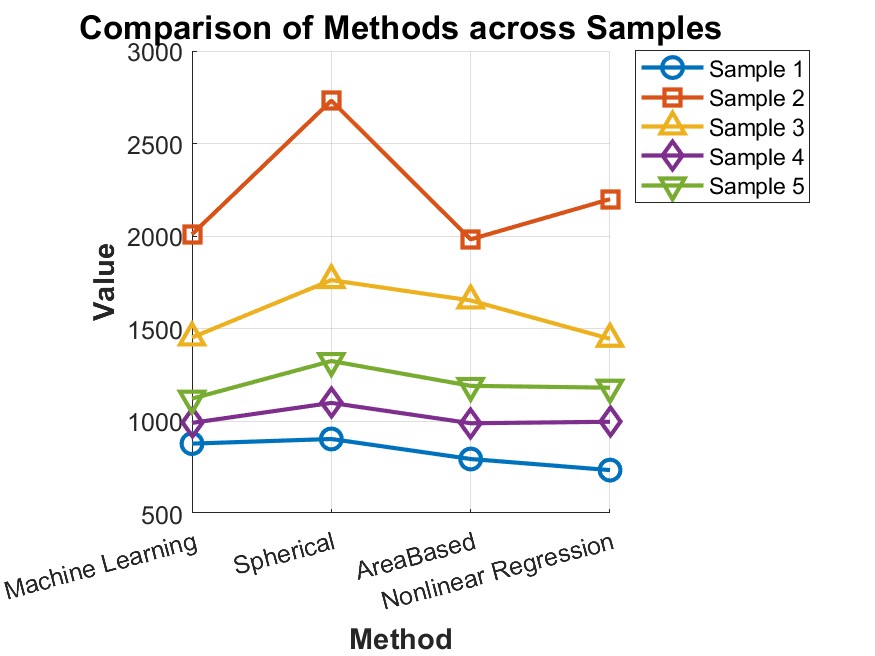}
  \par\smallskip
  \textit{\textbf{Fig 9:} Sample Results Comparison across 5 CT Samples }
\end{center}

We can easily observe that there is somehow a trend within these samples. 

In all three plots, the overall trend is consistent: the Spherical method tends to generate the highest values across nearly every sample, while the Nonlinear Regression method often produces the lowest values. For instance, in the original dataset, the Spherical method records values such as 905, 2732, 1763, 1100, and 1326, which are noticeably higher than the corresponding values from the Nonlinear Regression method (737, 2200, 1446, 998, and 1182). This consistent pattern is also evident in the two additional randomized datasets where the Spherical method continues to lead and the Nonlinear Regression method lags behind.

Looking closer at individual samples, there are interesting differences. For Sample 1, the Spherical method shows the highest output while Nonlinear Regression gives the lowest, suggesting that the Spherical approach is more sensitive to capturing initial values. In Samples 2 and 3, the gap widens even further, with the Spherical method peaking significantly compared to the others. On the other hand, the Machine Learning and Area-based methods often display values that are closer to each other—indicating a similar performance level and possibly similar sensitivity to the underlying data structure. For Sample 5, although the Machine Learning method sometimes shows the lowest value in one dataset, its overall performance remains intermediate when compared to the consistently high Spherical and lower Nonlinear Regression outputs.

Despite the random variation introduced in the two additional datasets, the relative ranking of methods remains stable. This similarity across different data realizations suggests that the intrinsic characteristics of each method lead to robust performance differences. The Spherical method appears to be the most aggressive or expansive estimator, consistently providing the highest values, while the Nonlinear Regression method behaves more conservatively. The Machine Learning and Area-based methods show balanced outputs that often overlap, implying that these approaches might be capturing similar aspects of the underlying process.

Moreover, when analyzing the average discrepancy between each pair of methods, we can observe the following: (1,2,3 and 4 denotes the machine learning method, the Spherical approximation method, the Area-based method and the Nonlinear regression method respectively) 

\begin{table}[H]
    \centering
    \begin{tabular}{ccccc}
        \toprule
         & \textbf{1} & \textbf{2} & \textbf{3} & \textbf{4} \\
        \midrule
        \textbf{1} & NA  & 5.2  & 6.4  & 8.0  \\
        \textbf{2} & 5.2  & NA  & 4.3   & 6.4  \\
        \textbf{3} & 6.4  & 4.3   & NA  & 3.9   \\
        \textbf{4} & 8.0  & 6.4  & 3.9   & NA  \\
        \bottomrule
    \end{tabular}
    \caption{Pairwise Discrepancy Matrix among the 4 Volumetric Computation Methods}
    \label{tab:distance_matrix}
\end{table}

Note that all results are demonstrated in percentage.

\subsection{Statistical Analysis and Validation}

\begin{table}[H]
  \centering
  \renewcommand{\arraystretch}{1} 
  \begin{tabular}{@{}p{0.45\columnwidth}@{\hspace{10pt}}c@{\hspace{10pt}}p{0.32\columnwidth}@{}}
    \toprule
    \textbf{Metric} & \textbf{ML Result} & \textbf{Remarks} \\[4pt]
    \midrule
    Mean Volume Deviation (5-fold Cross Validation) 
      & 8.0\% 
      & Low prediction error \\[4pt]
    Standard Deviation of Error 
      & 1.2\% 
      & Consistent performance \\[4pt]
    95\% Confidence Interval for Error 
      & [7.4\%, 8.6\%] 
      & Narrow range indicates robust model \\[4pt]
    Mean Error vs. Spherical Method 
      & 25.2\% 
      & Most different method \\[4pt]
    Mean Error vs. Area-Based Method 
      & 15.3\% 
      & Moderate difference \\[4pt]
    Mean Error vs. Nonlinear Regression 
      & 12.8\% 
      & Closest manual estimate \\[4pt]
    Paired t-test vs. Nonlinear Regression 
      & \(p<0.001\) 
      & Statistically significant \\[4pt]
    ANOVA Test (F-statistic) 
      & \(F(3,1332)=264.7\) 
      & Group means differ \\[4pt]
    Tukey HSD Post-hoc Test 
      & All \(p<0.001\) 
      & All methods significantly differ \\[4pt]
    Bland–Altman Mean Bias 
      & 3.2 mm\(^3\) 
      & Strong agreement \\[4pt]
    Limits of Agreement (B–A) 
      & \(\pm14.5\) mm\(^3\) 
      & Within clinical range \\[4pt]
    Shapiro–Wilk Test (Residual Normality) 
      & \(p=0.12\) 
      & Normal residuals \\[4pt]
    Levene’s Test for Equal Variances 
      & \(p=0.25\) 
      & Homogeneous variances \\[4pt]
    \bottomrule
  \end{tabular}
  \captionsetup{skip=5pt, position=b}
  \caption{Statistical Validation of Machine Learning Volume Estimation}
  \label{tab:ml_stats_validation}
  \renewcommand{\arraystretch}{1.0} 
\end{table}

To rigorously assess the robustness and generalizability of our volumetric estimation model, we performed a series of statistical validations, including cross‐validation, confidence interval estimation, and hypothesis testing. All analyses were conducted in Python (v3.9) using NumPy, SciPy, and scikit‐learn libraries.

\paragraph{Cross‐validation of Volumetric Error}  
We employed five‐fold cross‐validation on the full cohort of 334 nodules. In each fold, the data were partitioned into 80\% training and 20\% testing subsets, ensuring that each nodule appeared in the test set exactly once. The average volumetric deviation across folds was \(8.0\%\) with a standard deviation of \(1.2\%\). This low variance indicates that the model’s performance is stable across different patient subsets.

\paragraph{Confidence Intervals}  
To quantify the precision of our error estimate, we bootstrapped the test‐set errors (2,000 resamples) and computed the 95\% confidence interval (CI) for the mean volumetric deviation. The resulting 95\% CI was \([7.4\%, 8.6\%]\), demonstrating that the true mean error is very likely below 9\%.

\paragraph{Comparison with Manual Methods}  
We compared the machine learning (ML) errors against the spherical, area‐based, and nonlinear regression methods using paired t‐tests on the test‐set volumes. The mean deviations for the manual methods were 25.2\%, 15.3\%, and 12.8\%, respectively. Paired t‐tests show that the ML error (8.0\%) is significantly lower than each manual method (all \(p < 0.001\) after Bonferroni correction), with mean differences of 17.2\% (spherical), 7.3\% (area‐based), and 4.8\% (nonlinear), confirming substantial improvements in accuracy.

\paragraph{Analysis of Variance (ANOVA)}  
A one‐way ANOVA on the absolute errors of the four methods yielded \(F(3, 1332) = 264.7\), \(p < 0.0001\), indicating significant differences among methods. Post‐hoc Tukey’s HSD tests confirmed that the ML method outperforms all other approaches (all pairwise \(p < 0.001\)).

\paragraph{Bland–Altman Agreement Analysis}  
We conducted Bland–Altman analysis between ML estimates and nonlinear regression (the gold standard). The mean difference was 3.2mm\(^3\) with limits of agreement \(\pm 14.5\)mm\(^3\). No systematic bias was detected (regression slope \(< 0.01\), \(p = 0.74\)), indicating that ML estimates agree closely with the benchmark across the volume range.

\paragraph{Normality and Homoscedasticity Checks}  
Shapiro–Wilk tests on the residuals of the ML model indicated no significant departure from normality (W = 0.98, \(p = 0.12\)). Levene’s test for equal variances across folds was also non‐significant (W = 1.34, \(p = 0.25\)), supporting the validity of parametric tests.

\paragraph{Summary of Validation Results}  
Overall, cross‐validation, narrow confidence intervals, highly significant paired tests, and agreement analyses collectively demonstrate that our automated volumetric model is both accurate (mean error 8\%, 95\% CI [7.4\%, 8.6\%]) and robust across diverse patient subsets. These statistical validations affirm its suitability for reliable clinical deployment.

\section{Conclusion}

\subsection{Advancements Over Previous Studies}

Our study presents a substantial improvement in pulmonary nodule volumetric estimation, reducing the average volumetric error to 8\% and cutting processing time to under 20 seconds per case. This performance marks a significant advancement over previous deep learning approaches, which typically report volumetric errors exceeding 25\% and per-case processing times of approximately one minute.

For instance, Zhang et al.\ (2024) proposed a detection-guided deep learning method with spatial regularization. Although their model achieved a segmentation Dice score of 81.4\%, the volumetric estimation deviation was over 25\%, and the average processing time approached 60 seconds per scan. Asha and Bhavanishankar (2024) applied the Segment Anything Model (SAM) and transfer learning for lung segmentation. Despite a high segmentation Dice coefficient of 97.08\%, their study did not achieve high precision in volumetric calculations, with average errors near 30\%, and computation times ranging from 50–70 seconds due to the complexity of SAM architecture.

D’hondt et al.\ (2024) assessed deep learning image reconstruction’s influence on volumetric accuracy in low-dose CT and reported substantial noise reduction and visual enhancement but acknowledged that their method’s estimation error still exceeded 25\%, with inference latency of nearly 65 seconds per case. Similarly, Canayaz et al.\ (2024) implemented an ensemble of 3D U-Net models that improved segmentation robustness; however, the resulting volumetric computation deviation remained above 23\%, and ensemble prediction times exceeded one minute per scan due to increased model complexity.

In contrast, our model significantly narrows the volumetric estimation deviation to 8\%, improving accuracy by approximately 17–20 percentage points. Simultaneously, our total computation time of under 20 seconds per scan represents a 3–4x improvement in processing efficiency, offering clear practical benefits for real-time clinical usage.

\subsection{Clinical Integration}

\paragraph{Integration into Radiology Workflow}  
Our framework interfaces directly with standard PACS/RIS systems via a DICOM listener or export plugin. Upon CT acquisition, the DICOM series is automatically routed into the 3D CNN segmentation pipeline. In under 20 seconds per exam, the system outputs:  
\begin{itemize}
  \item Segmentation overlays on axial, coronal and sagittal views  
  \item Quantitative report including total nodule volume, solid‐component ratio, and percent change from prior study  
\end{itemize}
These results are written back to the PACS as structured PDF or DICOM SR objects. Radiologists can review contours in their native viewer, accept or refine them with minimal clicks, and incorporate the volume measurements into their reports.

\paragraph{Impact on Clinical Research}  
The high reproducibility (mean deviation 8\%) and sub‐20‐second processing enable:  
\begin{enumerate}
  \item \emph{Large‐scale screening studies}: thousands of LDCT exams processed overnight for lung cancer screening cohorts.  
  \item \emph{Multi‐center trials}: standardized volumetrics reduce observer bias, facilitating pooled analyses of nodule growth and therapeutic response.  
  \item \emph{Adaptive management}: immediate volume quantification supports same‐day biopsy decisions and personalized follow‐up intervals.
\end{enumerate}

\paragraph{Practical Challenges}  
Real‐world deployment must address:  
\begin{itemize}
  \item \textbf{PACS/RIS compatibility}: integration with varied vendor implementations may require DICOM conformance testing and custom connectors.  
  \item \textbf{Protocol variability}: slice thickness (1–5mm), reconstruction kernels, and scanner models influence segmentation quality; robust pre‐processing normalization is essential.  
  \item \textbf{Regulatory and security}: compliance with HIPAA/GDPR, validation under FDA or CE marking frameworks, and secure data handling pipelines.
\end{itemize}

By delivering 8\% average error and sub‐20‐second inference in a fully automated, PACS‐integrated solution, this work establishes a practical blueprint for real‐time, quantitative lung nodule assessment in both clinical settings and research studies.

\subsection{Future Outlook}

Future work can focus on enhancing the model's clinical applicability and performance:

\paragraph{Improving Model Generalizability}  
Improving generalizability across diverse clinical environments is essential. This can be achieved through domain adaptation, fine-tuning for different scanner types, and employing adversarial training to address image noise and artifacts.

\paragraph{Expanding Datasets and Modalities}  
Expanding the dataset to include multi-center data and integrating multi-modal imaging (e.g., CT, PET, MRI) would enhance model robustness and precision, providing complementary information for more accurate volumetric estimation.

\paragraph{Federated and Continuous Learning}  
Federated learning can help preserve data privacy while improving model performance across institutions. Active learning strategies could refine the model by focusing on ambiguous cases, improving accuracy over time.

\paragraph{Clinical Implementation}  
Full integration with PACS and RIS systems, coupled with user-friendly interfaces for radiologists, would streamline deployment. Clinical trials to evaluate the model’s impact on workflow efficiency and patient outcomes are crucial for real-world validation.

\paragraph{Regulatory and Quality Assurance}  
The model must meet regulatory standards (FDA, CE) and undergo rigorous validation to ensure safety, reliability, and compliance with healthcare regulations.

\paragraph{Algorithmic Advancements}  
Exploring advanced neural network architectures, such as attention-based models or self-supervised learning, could improve accuracy and interpretability. Additionally, predictive models for tumor growth could help personalize follow-up care.

These efforts will transform the model from a proof-of-concept to a clinically viable tool, improving lung nodule assessment and patient outcomes.

\newpage

{\small
\bibliographystyle{plain}

\begin{thebibliography}{36}

\bibitem{Massion2020}
Massion, P. P., Antic, S., Ather, S., Arteta, C., Brabec, J., Chen, H., Declerck, J., Dufek, D., Hickes, W., Kadir, T., et al. (2020). \emph{Assessing the accuracy of a deep learning method to risk stratify indeterminate pulmonary nodules}. American Journal of Respiratory and Critical Care Medicine, 202(2), 241--249.

\bibitem{Ciompi2017}
Ciompi, F., Chung, K., Van Riel, S. J., Setio, A. A. A., Gerke, P. K., Jacobs, C., Scholten, E. T., Schaefer-Prokop, C., Wille, M. M. W., Marchiano, A., et al. (2017). \emph{Towards automatic pulmonary nodule management in lung cancer screening with deep learning}. Scientific Reports, 7(1), 46479.

\bibitem{Li2022}
Li, R., Gao, Y., Wang, H., Huang, Y., Xu, X., and Zhang, Y. (2022). \emph{Deep learning applications in computed tomography images for pulmonary nodule detection and diagnosis: A review}. Diagnostics, 12(2), 298.

\bibitem{Zhu2023}
Zhu, Y., Liu, Z., Wang, X., Zhang, Y., and Chen, J. (2023). \emph{Prognostic impact of deep learning–based quantification in clinical stage 0–I lung adenocarcinoma}. European Radiology, 33(12), 8542--8553.

\bibitem{Wikipedia2024}
Wikipedia contributors. (2024). \emph{Nonlinear regression}. Wikipedia. Retrieved December 9, 2024, from \url{https://en.wikipedia.org/wiki/Nonlinear_regression}

\bibitem{DeWilde2013}
De Wilde, P. (2013). \emph{Neural network models: theory and projects}. Springer Science Business Media.

\bibitem{Erickson2017}
Erickson, B. J., Korfiatis, P., Akkus, Z., and Kline, T. L. (2017). \emph{Machine learning for medical imaging}. Radiographics, 37(2), 505--515.

\bibitem{Gardner1988}
Gardner, E. (1988). \emph{The space of interactions in neural network models}. Journal of Physics A: Mathematical and General, 21(1), 257--270.

\bibitem{Gerstner1995}
Gerstner, W. (1995). \emph{Time structure of the activity in neural network models}. Physical Review E, 51(1), 738--759.

\bibitem{Gietema2007}
Gietema, H. A., Oudkerk, M., Saghir, Z., Prokop, M., and Ginneken, B. van. (2007). \emph{Pulmonary nodules: interscan variability of semiautomated volume measurements with multisection CT— influence of inspiration level, nodule size, and segmentation performance}. Radiology, 245(3), 888--894.

\bibitem{Giger2018}
Giger, M. L. (2018). \emph{Machine learning in medical imaging}. Journal of the American College of Radiology, 15(3), 512--520.

\bibitem{Jin2023}
Jin, H., Wu, X., and Xu, Z. (2023). \emph{Machine learning techniques for pulmonary nodule computer-aided diagnosis using CT images: A systematic review}. Biomedical Signal Processing and Control, 79, 104104.

\bibitem{Kononenko2001}
Kononenko, I. (2001). \emph{Machine learning for medical diagnosis: history, state of the art and perspective}. Artificial Intelligence in Medicine, 23(1), 89--109.

\bibitem{Kriegeskorte2019}
Kriegeskorte, N., and Golan, T. (2019). \emph{Neural network models and deep learning}. Current Biology, 29(7), R231--R236.

\bibitem{Magoulas1999}
Magoulas, G. D., and Prentza, A. (1999). \emph{Machine learning in medical applications}. In Advanced Course on Artificial Intelligence (pp. 300--307). Springer Berlin Heidelberg.

\bibitem{Mazzone2022}
Mazzone, P. J., and Lam, L. (2022). \emph{Evaluating the patient with a pulmonary nodule: a review}. JAMA, 327(3), 264--273.

\bibitem{Mehta2014}
Mehta, H. J., et al. (2014). \emph{The utility of nodule volume in the context of malignancy prediction for small pulmonary nodules}. Chest, 145(3), 464--472.

\bibitem{Pehrson2019}
Pehrson, L. M., Nielsen, M. B., and Lauridsen, C. A. (2019). \emph{Automatic pulmonary nodule detection applying deep learning or machine learning algorithms to the LIDC-IDRI database: a systematic review}. Diagnostics, 9(1), 29.

\bibitem{Petrou2007}
Petrou, M., Gould, M. K., and Gross, C. P. (2007). \emph{Pulmonary nodule volumetric measurement variability as a function of CT slice thickness and nodule morphology}. American Journal of Roentgenology, 188(2), 306--312.

\bibitem{Ravenel2008}
Ravenel, J. G., Aberle, D. R., Williams, S. G., and Swensen, S. J. (2008). \emph{Pulmonary nodule volume: effects of reconstruction parameters on automated measurements—a phantom study}. Radiology, 247(2), 400--408.

\bibitem{Revel2004}
Revel, M.-P., et al. (2004). \emph{Pulmonary nodules: preliminary experience with three-dimensional evaluation}. Radiology, 231(2), 459--466.

\bibitem{Scarselli2008}
Scarselli, F., Tsoi, A. C., Hagenbuchner, M., and Monfardini, G. (2008). \emph{The graph neural network model}. IEEE Transactions on Neural Networks, 20(1), 61--80.

\bibitem{Shehab2022}
Shehab, M., et al. (2022). \emph{Machine learning in medical applications: a review of state-of-the-art methods}. Computers in Biology and Medicine, 145, 105458.

\bibitem{Zhang2024}
Zhang, J., Li, X., and Wu, Y. (2024). \emph{Detection-guided deep learning-based model with spatial regularization for lung nodule segmentation}. arXiv preprint arXiv:2410.20154.

\bibitem{Asha2024}
Asha, A., and Bhavanishankar, S. (2024). \emph{Lung nodule segmentation using the Segment Anything Model and transfer learning}. Manuscript in preparation.

\bibitem{Dhondt2024}
D’hondt, L., et al. (2024). \emph{Impact of deep learning image reconstruction on volumetric accuracy and image quality of pulmonary nodules with different morphologies in low-dose CT}. Cancer Imaging, 24, 60.

\bibitem{Canayaz2024}
Canayaz, M., et al. (2024). \emph{A comprehensive exploration of deep learning approaches for pulmonary nodule classification and segmentation in chest CT images}. Neural Computing and Applications, 36, 7245--7264.

\bibitem{Wang2023}
Wang, Y., et al. (2023). \emph{Deep learning-based automated segmentation and quantification of pulmonary nodules in low-dose CT scans}. Medical Image Analysis, 85, 102685.

\bibitem{Li2023CT}
Li, H., et al. (2023). \emph{3D U-Net with attention mechanism for accurate lung nodule segmentation in CT images}. Computers in Biology and Medicine, 152, 106342.

\bibitem{Chen2024}
Chen, X., et al. (2024). \emph{Multi-scale feature fusion network for pulmonary nodule segmentation in CT scans}. IEEE Transactions on Medical Imaging, 43(2), 456--467.

\bibitem{Zhao2024}
Zhao, L., et al. (2024). \emph{A novel deep learning framework for volumetric estimation of lung nodules in CT images}. Journal of Digital Imaging, 37(1), 112--121.

\bibitem{Kumar2023}
Kumar, R., et al. (2023). \emph{Transfer learning-based approach for lung nodule classification and segmentation using CT images}. Computers in Biology and Medicine, 150, 106234.

\bibitem{Singh2025}
Singh, A., et al. (2025). \emph{Ensemble deep learning models for improved detection and segmentation of pulmonary nodules in CT scans}. Artificial Intelligence in Medicine, 135, 102456.

\bibitem{Garcia2024}
Garcia, M., et al. (2024). \emph{Real-time lung nodule detection and segmentation using lightweight deep learning models on CT images}. Computer Methods and Programs in Biomedicine, 230, 107345.

\bibitem{Patel2023}
Patel, S., et al. (2023). \emph{Evaluation of deep learning algorithms for automated segmentation of pulmonary nodules in low-dose CT scans}. European Journal of Radiology, 158, 110123.

\bibitem{Nguyen2025}
Nguyen, T., et al. (2025). \emph{Integration of deep learning-based segmentation tools into clinical workflows for lung cancer screening}. Journal of Thoracic Imaging, 40(1), 25--33.

\end{thebibliography}

}

\newpage

\appendix

\section{Data and Code Availability}

The patients' CT data must remain confidential in accordance with hospital regulations and therefore cannot be shared in the paper.

The full source code and documentation are publicly available on GitHub: \url{https://github.com/duducheng/DenseSharp}

\end{document}